\documentclass[aps,prl,floatfix,twocolumn]{revtex4}

\usepackage{amsmath}
\usepackage{amssymb}
\usepackage{graphicx}
\usepackage{epstopdf}
\usepackage{color}
\usepackage{cancel}

\newcommand{\ket}[1]{|#1\rangle}		

\begin{document}

\title{Discrete, Tunable Color Entanglement}

\author{S.~Ramelow$^{\dagger,1,2}$, L.~Ratschbacher$^{\dagger,1}$, A.~Fedrizzi$^{1,2,3}$, N.~K.~Langford$^{1,2}$ and A.~Zeilinger$^{1,2}$}

\affiliation{
$^1$Institute for Quantum Optics and Quantum Information, Austrian Academy of Sciences, Boltzmanngasse 3, A-1090 Vienna, Austria \\
$^2$Faculty of Physics, University of Vienna, Boltzmanngasse 5, A-1090 Vienna, Austria \\
$^3$University of Queensland, Brisbane 4072, Australia }

\begin{abstract}
Although frequency multiplexing of information has revolutionized the field of classical communications, the color degree of freedom (DOF) has been used relatively little for quantum applications. We experimentally demonstrate a new hybrid quantum gate that transfers polarization entanglement of non-degenerate photons onto the color DOF. We create, for the first time, high quality, discretely color-entangled states (with energy bandgap up to 8.4 THz) without any filtering or postselection, and unambiguously verify and quantify the amount of entanglement (tangle, $0.611{\pm}0.009$) by reconstructing a restricted density matrix; we generate a range of maximally entangled states, including a set of mutually unbiased bases for an encoded qubit space. The technique can be generalized to transfer polarization entanglement onto other photonic DOFs, like orbital angular momentum.
\end{abstract}

\maketitle

Color, or frequency, is one of the most familiar degrees of freedom (DOFs) of light and has been routinely analyzed in spectroscopy for centuries.
However, although frequency multiplexing of information has had a profound impact on classical telecommunications, little work has aimed at exploiting the frequency DOF for quantum-based information technologies.  A key ingredient in many such technologies is discretely encoded entanglement, which has been extensively investigated for other optical degrees of freedom (e.g., \cite{kwiat_new_1995, KwiatPG1999a, KimT2006a, AlessandroSource, KwiatPG1993a, timeEntanglement, ThewRT2004a, RarityJG1990a, OAMentanglement, LangfordNK2004a}).  In contrast, discrete frequency entanglement has not yet been unambiguously demonstrated, despite potentially interesting applications such as enhanced clock synchronization beyond the classical limit~\cite{giovannetti_quantum-enhanced_2001,de_burgh_quantum_2005} and improved quantum communication in noisy channels~\cite{xiao_efficient_2008}.
Flying qubits encoded in tunable frequency bins would also be an ideal mediator between stationary qubits with different energy levels; e.g., very recently the state of two photons emitted by two separate Yb ions was projected onto a discrete frequency-entangled state, allowing the creation of entanglement and realization of teleportation between the ions~\cite{olmschenk_quantum_2009}.
Finally, the higher-dimensional Hilbert space accessible with the color DOF has known benefits for quantum communication~\cite{FujiwaraM2003a,WangC2005a} and quantum cryptography~\cite{BrussD2002a,CerfNJ2002a, SpekkensRW2001a}, and would also allow the exploration of fundamental questions about quantum mechanics~\cite{kochen-specker}. 

Continuous frequency entanglement between photon pairs arises naturally in spontaneous parametric down-conversion (SPDC) experiments as a consequence of energy conservation \cite{oumandel, raritybeating, KwiatPG1993a, alessandroarxive}.  It is often, however, much simpler to control and use entanglement between systems with discrete, well-separated basis states (cf.\ time-bin entanglement~\cite{timeEntanglement, ThewRT2004a}). A simple discrete color-entangled state would be $(\ket{\omega_1}\ket{\omega_2}{+}\ket{\omega_2}\ket{\omega_1})/\sqrt{2}$, where $\ket{\omega_j}$ represent single-photon states occupying discrete, well-separated frequency bins. Although such a state can be \emph{postselected} from broadband, continuous frequency entanglement (e.g., as in \cite{oumandel, raritybeating}), for most quantum applications it is necessary to explicitly create such a state without postselection. There have been some proposals and attempts to create discrete color entanglement in nonlinear waveguides~\cite{booth_counterpropagating_2002,ravaro_nonlinear_2005}.  To date, however, no experiment has been able to conclusively show the creation or quantitative measurement of discretely color-entangled photons.

Here we report the first experimental demonstration of genuine discretely color-entangled states, created without any spectral filtering or postselection.  We used a hybrid quantum gate (HQG), a gate that acts simultaneously on different DOFs, that can deterministically transfer polarization onto color entanglement and unambiguously verified and quantified this entanglement using non-classical interference.  We also demonstrated full control over the frequency separation and phase of the created states, while maintaining a high fidelity.

\begin{figure}[t]%
\includegraphics[width=\columnwidth]{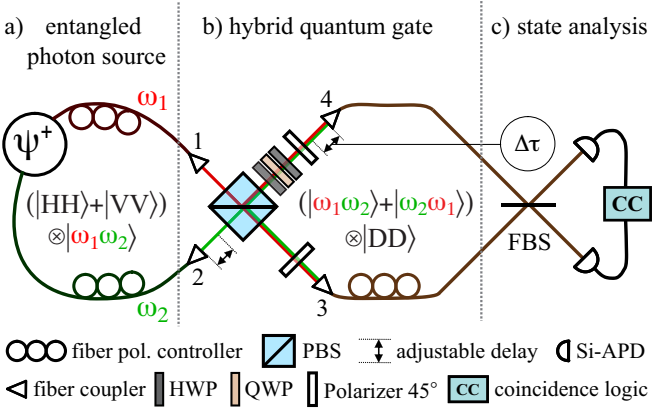}%
\caption{(Color online) Schematic of the experimental setup. (a) Source of polarization-entangled photon pairs with tunable central frequencies. (b) The hybrid quantum gate's polarizing beam splitter (PBS) maps the polarization entanglement onto the color degree of freedom. Subsequently projecting on diagonal (D) linear polarization with polarizers (POL) at 45$^{\circ}$ generates the discretely color entangled state. (c) The state is analyzed by two-photon interference at a fiber beamsplitter (FBS); Si-APD single-photon detectors and coincidence counting (CC) logic measure the coincidence rate as a function of temporal delay between modes.}%
\label{fig:figure1}%
\end{figure}

In our experiment (Fig.~\ref{fig:figure1}), a tunable source of polarisation entanglement based on continuous-wave SPDC~\cite{AlessandroSource} generates fibre-coupled photon pairs close to a pure state:
  \begin{equation}\label{eq:polentangledstate}
  \ket{\psi_{\rm in}}=(\alpha\ket{H}_1\ket{H}_2+e^{i\phi}\beta\ket{V}_1\ket{V}_2)\otimes\ket{\omega_1}_1\ket{\omega_2}_2,
  \end{equation}
where $\alpha^2+\beta^2=1$, $H$ and $V$ denote vertical and horizontal polarization, and $\omega_j$ is the central frequency of mode $j$. This notation neglects the spectral entanglement within the single-photon bandwidth,
which was much less than the photons' frequency separation, $\mu = \omega_1 {-} \omega_2$. 
By varying the temperature of the source's nonlinear (ppKTP) crystal, we continuously tuned the photon frequencies from degeneracy (809.6nm at 25.1$^\circ$C) to a maximum separation of 8.4 THz (18.3 nm) at 68.1$^\circ$C while maintaining high-quality polarization entanglement~\cite{AlessandroSource}. We controlled the polarization state with wave plates. 

Single mode fibers connect the source to the inputs of the hybrid gate depicted in Fig.~\ref{fig:figure1}(b). The PBS maps the state $\ket{\omega_1}_1$, depending on its polarization, to $\ket{\omega_1}_3$ ($H$) or $\ket{\omega_1}_4$ ($V$), and similarly for the state $\ket{\omega_2}_2$. This transfers the existing polarization entanglement onto color with the resulting \emph{hypoentangled}~\cite{nathanphd,hypoentangled} multi-DOF state:
 \begin{equation}\label{eq:hypocolorstate}
 \ket{\psi_{\rm hypo}}=\alpha\ket{H\omega_1}_3\ket{H\omega_2}_4+e^{i\phi}\beta\ket{V\omega_2}_3\ket{V\omega_1}_4.
 \end{equation}
To create the desired state, the frequency entanglement must then be decoupled from the polarization DOF.  This can be achieved deterministically by selectively rotating the polarization of one of the two frequencies (e.g., using dual-wavelength wave plates).  For simplicity, we instead chose to erase the polarization information probabilistically by projecting both photons onto diagonal polarization using polarizers at 45$^\circ$. We erased temporal distinguishability between input photons by translating fibre coupler 2 to maximise the non-classical interference visibility at the PBS for degenerate photons. Finally, we compensated for unwanted birefringent effects of the PBS using wave plates in one arm. The gate output is then:
  \begin{equation}\label{eq:purecolorstate}
  \ket{\psi_{\rm out}}=\alpha\ket{\omega_1}_3\ket{\omega_2}_4+e^{i\phi}\beta\ket{\omega_2}_3\ket{\omega_1}_4.
  \end{equation}
The parameters defining this state can be set by preparing an appropriate polarization input state (Eq.~\ref{eq:polentangledstate}).

\begin{figure}[t]
 \includegraphics[width=\columnwidth]{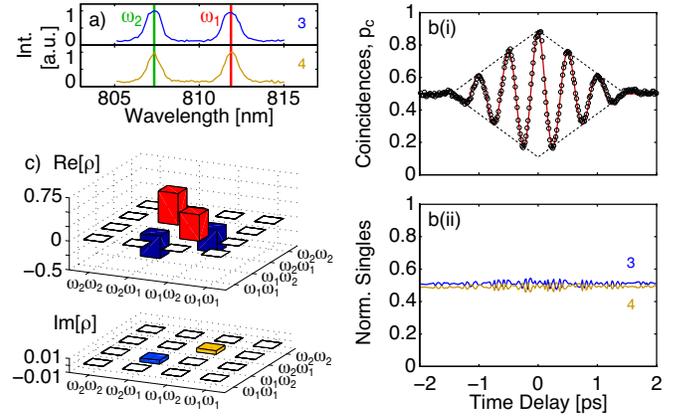}%
 \caption{(Color online) Analysis of the discretely color-entangled state. a) Single-photon spectra for modes 3 and 4; frequency separation is 2.1 THz (4.6 nm). The observed width of each bin is limited by the single-photon spectrometer. b) Normalized (i) coincidence and (ii) singles count rates as a function of delay in mode 4. The solid line in (i) is a fit of Eq.~\ref{eq:beatingfunction} to determine $V$ and the phase $\phi$. c) The estimated restricted density matrix: target-state fidelity, $0.891{\pm}0.003$; tangle, $0.611{\pm}0.009$; and purity, $0.801{\pm}0.004$.}
 \label{fig:figure2}
 \end{figure}

To explore the performance of the hybrid gate, we first injected photon pairs close to the polarization state $(\ket{H}_1\ket{H}_2{-}\ket{V}_1\ket{V}_2)/\sqrt{2}$ with individual wavelengths 811.9 nm and 807.3 nm.  The gate should then ideally produce the discrete, anticorrelated color-entangled state: $\ket{\psi}=(\ket{\omega_1}_3\ket{\omega_2}_4{-}\ket{\omega_2}_3\ket{\omega_1}_4)/\sqrt{2}$.  Figure~\ref{fig:figure2}a) shows the unfiltered single-photon spectra of the two output modes, illustrating that each photon is measured at either $\omega_1$ or $\omega_2$. This reflects a curious feature of discretely colour-entangled states, that individual photons have no well-defined color and no photon is ever observed at ``mean-value'' frequency. This feature clearly distinguishes our experiment from the continuous frequency entanglement studied in earlier work~\cite{oumandel, raritybeating, KwiatPG1993a}.

Because the detuning, $\mu=4.6$ nm, is much larger than the FWHM bandwidth of the individual color modes of 0.66 nm (0.30 THz; defined by the 10 mm nonlinear crystal), the two modes are truly orthogonal, making them good logical states for a frequency-bin qubit. This orthogonality also means that color anticorrelations are strictly enforced by energy conservation, because a single down-conversion event cannot produce two photons in the same frequency bin. We confirmed this by directly measuring the gate output in the frequency-bin computational basis (i.e.\ with bandpass filters in each arm tuned to $\omega_1$ or $\omega_2$). We observed strong, comparable coincidence rates for the two ``anticorrelated'' basis states ($10882\pm104$ and $9068\pm95$ in 30s for $\ket{\omega_1}_3\ket{\omega_2}_4$ and $\ket{\omega_2}_3\ket{\omega_1}_4$, resp.), and no coincidences for the same-frequency states ($\ket{\omega_1}_3\ket{\omega_1}_4$ and $\ket{\omega_2}_3\ket{\omega_2}_4$) to within error bars determined by the filters' finite extinction ratios.

To demonstrate that the color state was not only anticorrelated but genuinely entangled, we used nonclassical two-photon interference~\cite{HongCK1987a}, overlapping the photons at a 50:50 fibre beam splitter (FBS) (Fig.~\ref{fig:figure1}c) and varying their relative arrival time by translating fibre coupler 4 while observing the output coincidences. The results in Fig.~\ref{fig:figure2}b) show high-visibility sinusoidal oscillations (frequency $\mu$) within a triangular envelope caused by the \emph{unfiltered} ``sinc-squared'' spectral distribution of the source~\cite{AlessandroSource,triangledip}. At the central delay, the normalised coincidence probability reaches up to $0.881\pm0.007$, far above ($>50\sigma$) the baseline level of $0.5$. This antibunching is an unambiguous signature for antisymmetric entanglement \cite{MattleK1996a,BStheory,alessandroarxive} and, in conjunction with the previous measurements, conclusively demonstrates that our discrete color state is strongly entangled. As expected, the single-photon detection rates (Fig.~\ref{fig:figure2}b) exhibit negligible interference effects. 
It is important to note that this signature is similar to those observed in earlier ``spatial quantum beating'' experiments~\cite{oumandel, raritybeating}. As demonstrated by Kaltenbaek et al.~\cite{KaltenbaekR2009a}, however, observing this signature in different contexts does not necessarily lead to the same conclusions. In previous experiments, the observed signal was only \emph{postselected} from broadband continuous frequency entanglement and at no point could the quantum state of the photons emitted by the source be described solely as a discretely colour-entangled state, uncoupled from other DOFs. By contrast, our measurements do not rely on any spectral postselection, but as supported by the single-photon spectra, they result directly from unfiltered discrete color entanglement.

We now show how we can combine the above measurements to estimate a restricted density matrix in colour space. We first recall that energy conservation in the SPDC pair source and during photon propagation constrains the state to the two-dimensional anticorrelated subspace of the two-qubit color space (before and after the gate). This is a physical constraint, validated by the measurements in the computational basis. The complete density matrix within this subspace can be written (in the computational basis, $\{ \ket{\omega_1}_3\ket{\omega_1}_4, \ket{\omega_1}_3\ket{\omega_2}_4, \ket{\omega_2}_3\ket{\omega_1}_4, \ket{\omega_2}_3\ket{\omega_2}_4\}$):
 \begin{equation}\label{eq:colorentangledstate}
  \rho= \left( \;
  \begin{matrix}
	0 \quad & \quad 0 \quad &  \quad 0 \quad & \quad 0 \\
	0 \quad & p & \frac{V}{2}\:e^{-i\phi} & \quad 0 \\
	0 \quad & \frac{V}{2}\:e^{i\phi} &  1-p & \quad 0 \\
	0 \quad & 0 &  0 & \quad 0
  \end{matrix}
  \; \right)
  \end{equation}
with real parameters that obey the physicality constraints: $0\leq p\leq 1$ and $0\leq \frac{V}{2}\leq \sqrt{p(1-p)}$. Any detection events outside this subspace arise from higher-order emissions and accidental coincidences, and also lie outside the full two-qubit space.  Our computational basis measurements showed that these vanished to within error bars, and we directly calculated the balance parameter, $p=0.546{\pm}0.004$ (using Poissonian errors). We estimated the remaining parameters by fitting them to the nonclassical interference signal. For the above density matrix, given the source's spectral properties, we analytically calculated the expected interference probability, $p_c$, to be (following~\cite{alessandroarxive}):
  \begin{equation}
  p_c(\tau) = \begin{array}{c}
  \frac{1}{2}-\frac{V}{2}\cos(\mu\tau{+}\phi) \, (1{-}\left|\frac{2\tau}{\tau_c}\right|)  \quad \text{for }|\tau|<\frac{\tau_c}{2},
  \end{array}
  \label{eq:beatingfunction}
  \end{equation}
where the coherence time $\tau_c$ is the base-to-base envelope width, related to the single-photon frequency bandwidth via $\Delta f_{FWHM} = 0.885/\tau_c \sim 0.3$~THz. The missing elements V and $\phi$ can be identified as the visibility and phase of the oscillating signal and can therefore be estimated using curve fitting (for this state, $V= 0.782 {\pm} 0.006$ and $\phi = 179.2 {\pm} 0.4^\circ$).  The resulting density matrix (Fig.~\ref{fig:figure2}c) is strongly entangled, with a target-state fidelity of $0.891{\pm}0.003$, tangle~\cite{CoffmanV2000a} of $0.611{\pm}0.009$, and purity of $0.801{\pm}0.004$ (error bars include Poissonian and fitting errors). This is the first quantitative measurement of the entanglement of any color-entangled state.

\begin{figure}[t]%
\includegraphics[width=\columnwidth]{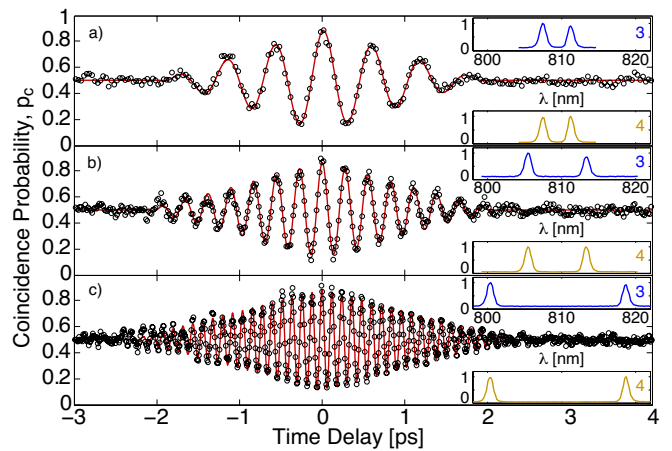}%
\caption{(Color online) Two-photon interference for color-entangled states with three different frequency separations (and corresponding crystal temperatures): a) 1.7 THz (3.8 nm), 33.7$^\circ$C; b) 3.6 THz (7.9 nm), 43.7$^\circ$C; and c) 8.4 THz (18.3 nm), 68.1$^\circ$C.  Solid lines show the curve fit according to Eq.~\ref{eq:beatingfunction} with $V$, $\mu$ and $\phi$ as fitting parameters.  The insets show the measured single-photon spectra for both modes of each state.}
\label{fig:figure3}
\end{figure}

Several error sources in our experiment contributed cumulatively to unwanted photon distinguishability in the final colour state and reduced the measured entanglement, including: imperfect input polarization states, imperfect mode matching and residual polarization misalignment at the PBS, the finite PBS extinction ratio, and a slightly asymmetric FBS splitting.  Accidental coincidence counts caused by detector dark counts and higher-order SPDC contributions were negligible.

To illustrate the flexibility of the hybrid gate, we analysed a series of output states for different frequency detunings $\mu$ and phases $\phi$. We first tuned $\mu$ by varying the crystal temperature in the source, and the results (Fig.~\ref{fig:figure3}) agree well with Eq.~(\ref{eq:beatingfunction}). The source enabled us to reach a detuning of 18.3 nm (8.4 THz), about 30 times the individual color-bin bandwidths. The detunings estimated from curve fitting matched the single-photon spectra.

\begin{figure}[t]
\includegraphics[width=\columnwidth]{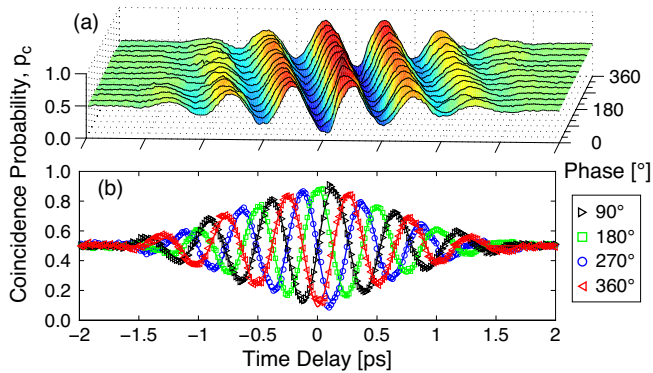}%
\caption{(Color online) (a) Coincidence probabilities after the FBS as a function of the delay for 13 different, close to maximally entangled discrete color states. The phase of the oscillation pattern is proportional to the phase of the original polarization-entangled state. (c) Four close to maximally entangled discrete color states that represent two unbiased bases.}
\label{fig:figure4}%
\end{figure}

We next prepared discrete color states of the form $(\ket{\omega_1}_3\ket{\omega_2}_4{+}e^{i\phi}\ket{\omega_2}_3\ket{\omega_1}_4)/\sqrt{2}$ with varying phase ($\phi=0^\circ,30^\circ,...,360^\circ$) (Fig.~\ref{fig:figure4}). The measured states display an average target-state fidelity of $0.90\pm0.01$, a tangle of $0.63\pm0.03$, and a purity of $0.82\pm0.02$, demonstrating that the hybrid gate accurately preserves quantum information stored in the original polarisation state. Note that, together with the product states $\ket{\omega_1}_3\ket{\omega_2}_4$ and $\ket{\omega_2}_3\ket{\omega_1}_4$, the entangled states with phase $0^\circ$, $90^\circ$, $180^\circ$ and $270^\circ$ constitute a full set of qubit mutually unbiased bases. This illustrates the states' potential usefulness for quantum protocols such as quantum cryptography.

In this paper, we have for the first time conclusively demonstrated the creation, control and characterisation of high-quality, discretely color-entangled states, prepared without any spectral filtering or postselection using a hybrid quantum gate. We performed the first quantitative measurement of color entanglement using a novel technique for characterising the two-qubit color state within a restricted, antisymmetric subspace defined by energy conservation.  Our hybrid gate can in fact be used to transfer polarization entanglement onto any desired photonic DOF ($\xi$), by preparing the input $\ket{\psi}_{\rm pol}\otimes\ket{\xi_1,\xi_2}$ and by appropriately erasing the polarization information after the PBS.  Because the preparation of high-quality polarization states can be much easier than in other photonic DOFs, this gate represents a valuable tool for quantum information processing tasks in those DOFs. Our work also has important implications for the development of quantum memories and repeaters, because color-encoded information could provide a natural interface between flying and stationary qubits (such as single ions, atoms or atom ensembles) where information is encoded in different energy levels. Indeed, by inverting the procedure from~\cite{olmschenk_quantum_2009}, one could potentially entangle distant ions directly by letting them absorb a photon pair with the appropriate discrete color entanglement.
Finally, we note that non-postselected, discretely color-entangled states could also be extracted from sources of continuous spectral entanglement (such as traditional SPDC) using custom-designed multi-band-pass filters.  Although this approach would not be easily tunable and efficient as ours is, it would allow access to higher-dimensional entangled states in the color DOF.

During preparation of this paper, some related work was published by X.\ Li \emph{et al.}~\cite{LiX2009a}. The authors report the creation of frequency-entangled photons using four-wave mixing in nonlinear fibres, but, as in \cite{oumandel, raritybeating}, used two narrow-band filters to postselect the desired state from a broader spectral distribution and color entanglement could not be demonstrated unambiguously.  

We would like to thank Thomas Jennewein and Bibiane Blauensteiner for useful ideas and support. This  work  has  been  supported  by  the  FWF  within SFB 015 P06, P20 and CoQuS (W1210), the European Commision Project QAP (No.\ 015846), and the DTO-funded U.S.\ Army Research Office QCCM program.


\end{document}